\documentclass[pra,twocolumn,showpacs]{revtex4}
\usepackage{graphicx}
\begin{document}

\title{Quantum advantages in classically defined tasks}

\author{N.  Aharon and L. Vaidman }
\affiliation{ School of Physics and Astronomy\\
Raymond and Beverly Sackler Faculty of Exact Sciences \\
Tel-Aviv University, Tel-Aviv 69978, Israel}

\date{\today}

\begin{abstract}
We analyze classically defined games for which a quantum team has an
advantage over any classical team. The quantum team has a clear
advantage in games in which the players of each team are separated
in space and the quantum team can use unusually strong correlations
of the Einstein-Podolsky-Rosen (EPR) type. We present an example of
a classically defined game played at {\em one} location for which
quantum players have a real advantage.
\end{abstract}

\pacs{03.67.-a, 03.65.Ta, 03.65.Ud}

\maketitle


Quantum information research shows how quantum devices can
outperform devices working on the basis of classical physics for
certain communication and computational tasks. One of the clear ways
to compare between the strength of quantum and classical methods is
to consider advantages of a quantum team playing games against a
classical team. Many papers give the impression that for nearly all
games ``quantum strategies'' are advantageous compared to classical
strategies \cite{Eisert,Benj,Meyer,Lee, Li,Fli,Nature,Iqbal}. Van
Enk and Pike \cite{VanEnkCF,VanEnkQG}, however, have pointed out
that quantized classical games differ as games from thier original
classical counterparts, and that in many cases quantum players
cannot win against classical players as long as the rules of the
game are unchanged. We find that in these games it is also important
to analyze the role of decoherence resulting from actions of
classical players. This decoherence frequently eliminates the
advantage of quantum players.

Quantum objects and strategies can  be useful in many contexts. In
quantum cryptography applications, quantum devices can replace the
third trusted party needed for some games (e.g. quantum gambling
\cite{Goldenberg}). In numerous cases where constraints on resources
are involved, a quantum team with $N$ qubits is much more efficient
than a classical team with $N$ bits \cite{Ga,VM}, although it is not
really a fair  comparison. This raises the question: In which games,
under  {\em equal} natural conditions, does a quantum team win
against a classical team?

We can define a particular game as a competition for factoring large
numbers. A quantum player, using Shor's algorithm \cite{Shor} should
win against a classical player by performing this task faster.
However, it is not clear when a quantum computer which outperforms a
classical computer will be built and, moreover, we  do not have
proof that a classical efficient algorithm does not exist.

Even less clear is that a quantum team can win in a competition of
minimum time for finding the answer to the Deutsh-Jozsa problem
\cite{Deutsch}, as Meyer suggests \cite{MeyerReply}. In this game a
black box is given which calculates a function for various inputs.
Even if we assume unlimited technological power of the quantum team
outside the black box, we cannot be sure that inside the box the
coherence needed for quantum computation is preserved. We can
imagine a quantum box which preserves coherence and which  can also
serve as a classical box for each possible input, but it is not a
particularly interesting observation that a classical team cannot
operate quantum devices efficiently. The question raised in this
regard is: Can a quantum team outperform a classical team in
classically defined tasks?

There is a well-known class of games in which a quantum team  with
good quantum devices  can unambiguously outperform any classical
team. We call them Einstein-Podolsky-Rosen (EPR) games since the
advantage of the quantum team is based on the use of entangled
systems exhibiting EPR correlations  which are stronger than any
possible classical correlations. Other names associated with these
games are {\it pseudo telepathy} \cite{Brassard,Cab} and Bell games
\cite{Tsirelson}.

 In EPR type
games each team has two or more players at separate locations. There
is a known set of questions the players can receive, a known set of
possible answers, and a payoff table for these answers. The players
are not allowed to communicate once the game begins (so they do not
know which questions the other team members were asked) but they are
allowed to communicate beforehand and share any physical devices
which might help them. The way to enforce the rule that these
devices must not allow the players to communicate during the game,
is to have the players make their moves before light signals
signifying the other players being given their questions  can
arrive. There are many examples of EPR games
\cite{DiViPe,KH00,Ca,SvD}. Conceptually, the simplest and clearest
EPR game is the one based on the Greenberger-Horne-Zeilinger proof
of nonlocality \cite{GHZ,Mermin,Vaidman GHZ}. Note also games based
on the Zeno-type proof of Bell inequalities
\cite{Hardy,VaidmanBell}.

Using a key distribution protocol \cite{BB84}, one can construct a
(rather artificial) game with players at two separate locations in
which teams equipped with quantum devices can have an advantage even
without  entanglement. The task is to transmit a message from one
laboratory to another through an optical fiber. One team has players
at two laboratories, while the second team  has an access to the
fiber. The second team gains points for correct guesses of  the
transmitted messages. The first team gains points when it correctly
catches the attempts of eavesdropping, but loses points if it
announces eavesdropping when the opponent has not touched the fiber.
Teams with quantum technology will have an advantage when the
allowed prior shared information is less than one time pad, but
enough to run the quantum protocol \cite{BB84}.

The question we want to analyze here is: Are there games played at
{\it one} location for which  quantum players have an advantage? An
important candidate for such a game is Meyer's coin flipping problem
\cite{Meyer}. A coin is placed heads up. Alice, in her first move,
can either flip or not flip the coin. Bob, in his subsequent move,
can also  either flip or not flip the coin, but he is not allowed to
see the state of the coin. Alice gets another turn in which she can
again either flip or not flip the coin without looking at its state.
She wins if the final state of the coin is heads up.

Classically, each player has maximally a 50\% chance of winning.
Meyer claims that using quantum mechanics, Alice can reach a 100\%
chance of winning. Meyer's proposal is that Alice, in her first
move, should put the coin in the superposition
\begin{equation}\label{1}
    {1\over
\sqrt 2} (|head\rangle + |tail\rangle).
\end{equation}
Then, whatever Bob does, either flip or not flip, the state of the
coin remains unchanged, and Alice in her last move can rotate the
(quantum) state back to $|head\rangle $.

Van Enk \cite{VanEnkCF} analyzed a particular realization of Meyer's
proposal in which the sides of the coin were represented as photon
polarization states and showed that quantum rotation to the
superposition (1) is actually a classical rotation of the
polarization that Alice, even without quantum capabilities, can
perform. So, van Enk concluded that even classically, Alice can
reach a 100\% chance of winning.

Discussing Meyer's proposal requires  specifying  its actual
realization. Van Enk mentioned that when we consider an actual coin,
the classical analog of Alice's quantum action is putting the coin
on its edge, which also yields a 100\% chance of winning. However,
this is clearly a different game because the set of allowed moves is
enlarged. Note that a coin standing on its edge is not described by
the state (1).

In Meyers's game with a real coin and original rules,  neither the
classical nor the quantum Alice can really always win. Indeed, even
if Alice, equipped with unlimited quantum technology, is capable of
creating the state (1), classical Bob will not leave it unchanged
after his turn. He is not supposed to perform a careful, precise
quantum experiment. Clearly, when Bob takes the coin in his hand,
its quantum state will decohere and Alice will not be able to rotate
it back to the state $|head\rangle$.

Formally, Meyer's idea, in which a quantum player puts the system in
a state that  moves of the classical player do not change, provides
an advantage for the quantum player. However, we are not aware of
any natural implementation of it as a real game in which a quantum
team, even with unlimited technological power, will have an
advantage over a classical team. The game involves a classical
player, and he invariably causes decoherence of the quantum state,
thereby eliminating the advantage of the quantum player.

We claim that there is at least one game,  played at one location,
in which  quantum players can get better results than  classical
players. This is the game based on the {\it three-box paradox}
\cite{Aharonov}. Contrary to EPR games which do not involve quantum
objects, but where the quantum player uses a quantum device to get
the right advice for a classical move, in this game, as in Meyer's
game, the object we play with is itself quantum. And, as in Meyer's
game, Bob does not see that the object he is playing with is a
quantum one. The difference is that, according to our game's rules,
Bob's actions do not cause decoherence which ultimately spoils
Alice's quantum moves.

Our game is a three stage game in which each player makes his moves
privately. Alice begins the game by preparing a single particle
which she places inside two boxes or any other place other than the
two boxes. The particle  can be prepared in any possible state
chosen by Alice. Bob, who has no information about the chosen state
of the system, can make one of two possible moves, either look for
the particle in box $A$ or look for the particle in box $B$. To
avoid any possible cheating by Alice, Bob can occasionally, instead
of his legitimate move, open two boxes to make sure that Alice does
not use two particles. Alice is not allowed to see Bob's move, but
there is a third trusted party which observes Bob's action and which
can see if Bob finds the particle. Bob's objective is to leave no
trace of his action, so he tries to leave the box exactly as it was
before. He is not allowed to touch the box which he chooses not to
open. Then Alice, in her turn, gets access to the boxes and can
perform any measurement she wants. She then has the option of either
canceling or accepting this trial of the game. She wins if Bob finds
the particle. Alice's objective is to maximize the probability of
the trails she does not cancel in which Bob finds the particle.

It is clear that  Alice who can use only classical objects cannot
obtain more than 50\% chance of winning. It seems that placing the
particle outside the two boxes can only reduce Bob's chances of
finding the particle and consequently  Alice's chances of winning.
Placing the particle in one of the boxes $A$ or $B$ leads to a 50\%
chance of Bob finding the particle. Alice's last move seems useless;
Bob's finding or not finding the particle does not change the
system, so her allowed measurement cannot help. She gets no
information about those cases when it is in her interest to cancel
the game trial.

However, Alice equipped with quantum devices can reach 100\% chance
of winning. She prepares the particle in a quantum state which is a
superposition of being in three boxes $A$, $B$, and $C$. The boxes
$A$ and $B$ are the ones Bob plays with. The third box she keeps for
herself; Bob need not know about it. The state is:
\begin{eqnarray}
\left|\psi\right\rangle  & = &
\frac{1}{\sqrt{3}}\left(\left|A\right\rangle +\left|B\right\rangle
+\left|C\right\rangle \right),\label{eq:3box_pre}
\end{eqnarray}
 where the  states $\left|A\right\rangle $, $\left|B\right\rangle $,
and $\left|C\right\rangle $ denote the particle being in box $A$,
$B$, and $C$, respectively.

Now, Bob opens either box $A$ or box $B$. He has a chance of one
third to find the particle in the box. Let us assume he opens box
$A$. (The game is symmetric with respect to the choices of $A$ and
$B$.) If Bob finds the particle in the box, its quantum state
becomes
\begin{equation}\label{A}
    | \psi_{FIND}\rangle = \left|A\right\rangle.
\end{equation}
If he does not find the particle in the box, its quantum state
becomes
\begin{equation}\label{BC}
    | \psi_{NOT~FIND}\rangle  = \frac{1}{\sqrt{2}}\left(\left|B\right\rangle
+\left|C\right\rangle \right).
\end{equation}
Since Bob is not allowed to touch the other box, i.e. box $B$, the
final quantum state in this case is exactly (\ref{BC}). We assume
that Alice's technological abilities are sufficient to build robust
boxes which, if untouched, keep the quantum state of the particle
inside it undisturbed. Bob's action with box $A$, even if he is not
careful, will not cause a change in the state (\ref{BC}). Bob tries
to leave no trace of his action, but if  he finds the particle in
$A$ and he is not careful, he might disturb the quantum state
(\ref{A}).

Alice, in her turn, makes  a projective measurement of the particle
on the state
\begin{eqnarray}
\left|\phi\right\rangle   =
\frac{1}{\sqrt{3}}\left(\left|A\right\rangle +\left|B\right\rangle
-\left|C\right\rangle \right).\label{eq:3box_post}
\end{eqnarray}
If she finds the state, she accepts the game trial, and if she does
not, she cancels it.

Now we see that Alice cannot lose:
\begin{equation}\label{loose}
\left\langle \psi_{NOT~FIND}\right| \phi\rangle =0,
\end{equation}
so the probability of Alice finding this particular state if Bob did
not find the particle is zero. And this is not sensitive to Bob's
action, provided he follows the rules. If Bob does find the particle
and the final state is (\ref{A}), we obtain:
\begin{equation}
 \label{win} \left\langle \psi_{FIND}\right| \phi\rangle
 =\frac{1}{\sqrt{3}}.
\end{equation}
Thus, Alice will accept the game with a probability of one third.
This probability becomes smaller if Bob is not careful and disturbs
the state of the particle in box $A$.  Alice declares ``game on''
only in the trials she wins, and never when she would lose.

We have shown that apart from games played in separate locations, in
which the EPR correlations give advantage to a quantum team, there
are classically defined games in one location  in which a classical
player unaware of quantum mechanics should not suspect anything
strange except for the unexplained fact that he loses. The essence
of the quantum team's advantage  here is that, whereas in classical
physics during ``an observation of a particle'' either we find it or
we don't, we do not change the state of the particle, in quantum
mechanics ``observation of a particle'' does change its state,
provided that the particle started in a superposition. (Compare this
with Meyer's example in which the action always changes
``classical'' states, and does not change the superposition state.)

Note that it is possible to find EPR correlations in our system.
Indeed, there is an entanglement between boxes. It is possible to
devise local experiments at different boxes showing violation of
Bell inequality \cite{NLAV}. However, in our game we do not have a
team of players each addressing a particular box, so this
entanglement is not the source of the advantage of the quantum
player. The  locality aspect of our game, i.e. that  the three boxes
are not at the same place, is crucial for the issue of decoherence.
Opening one box does not disturb  the relative phase between parts
of the quantum wave in other boxes.

One might gain an additional insight from  viewing our game in the
framework of the two-state vector approach \cite{AV90}. The essence
of the quantum advantage in this picture is that while the state of
a classical system at a particular time yields everything one can
know about this system given a known environment, in quantum
mechanics, future measurements might add information about the
present of a quantum system even if everything about the past is
known. This is why quantum Alice can benefit from her measurement
after Bob's observation.

Although there are real experiments testing these quantum
predictions \cite{Resch}, and there are demonstrations of  other
games \cite{Du}, today's technology does not yet enable one to win
games using quantum devices \cite{VaidmanBell}. It seems, however,
that we are not very far from this stage in technological
development.

Finally, we hope that our analysis of  transforming the three box
paradox into a game in which a quantum team wins against any
classical team will put an end to the controversy about the
classical analogy of the three-box paradox \cite{Kirkpatrick
2003a,Leifer,Ravon,Kirkpatrick 2007}. In all proposed classical
``analogies'' of the three-box paradox the intermediate measurement
changes the state of the system, while observation of a classical
particle in a box, does not.

We thank anonymous referees for helpful comments. This work has been
supported in part by the European Commission under the Integrated
Project Qubit Applications (QAP) funded by the IST directorate as
Contract Number 015848 and by grant 990/06 of the Israel Science
Foundation.

\end{document}